\documentclass[letterpaper, draftclsnofoot, onecolumn]{IEEEconf}
\usepackage{soul}
\usepackage[normalem]{ulem}

\usepackage[hidelinks]{hyperref}
\usepackage{bookmark}
\usepackage[latin1]{inputenc}
\usepackage[english]{babel}

\usepackage[dvipsnames]{xcolor}
\usepackage{amsfonts,amssymb,amsmath,color}
\usepackage[noadjust]{cite}
\usepackage{placeins}
\usepackage{graphicx}
\usepackage{algorithm}
\usepackage[]{algpseudocode}

\usepackage{tikz}
\usetikzlibrary{shapes,arrows}

\newcommand{\EE}{\mathcal{E}} 
\newcommand{\GG}{\mathcal{G}}

\newcommand{\lexmin}{\mathop{\rm lex\text{-}min}}

\newcommand{\subj}{\textnormal{subj. to}}

\newcommand{\nbrs}{\mathcal{N}}

\newcommand{\real}{{\mathbb{R}}}

\newcommand{\integer}{{\mathbb{Z}}}
\newcommand{\conv}[1]{\textnormal{conv}(#1)}

\newcommand{\until}[1]{\{1,\ldots,#1\}} 

\newcommand{\convXi}{\textnormal{conv}(X_i)}

\newcommand\oprocendsymbol{\hbox{$\square$}}
\newcommand\oprocend{\relax\ifmmode\else\unskip\hfill\fi\oprocendsymbol}

\newtheorem{theorem}{Theorem}[section]
\newtheorem{proposition}[theorem]{Proposition}

 \newtheorem{lemma}[theorem]{Lemma}
\newtheorem{remark}[theorem]{Remark}

\newtheorem{assumption}[theorem]{Assumption}

\newcommand{\bx}{\mathbf{x}}
\newcommand{\tx}{\mathbf{x}^{\textsc{L}}}
\newcommand{\hx}{\widehat{\mathbf{z}}}

\newcommand{\hz}{\widehat{\mathbf{z}}}

\newcommand{\bz}{\mathbf{z}}

\newcommand{\by}{\mathbf{y}}

\newcommand{\0}{\mathbf{0}}
\newcommand{\1}{\mathbf{1}}

\newcommand{\bmu}{\boldsymbol{\mu}}

\newcommand{\smallsum}{\textstyle\sum\limits}

\newcommand{\LP}{\textsc{lp}}
\newcommand{\MILP}{\textsc{milp}}

\newcommand{\INT}{\textsc{int}}
\newcommand{\ASY}{\textsc{asy}}
\renewcommand{\restriction}{\boldsymbol{\sigma}}
\newcommand{\ASYrestriction}{\sigma^\textsc{ASY}}

\newcommand{\bL}{\mathbf{L}}

\makeatletter
\newcommand{\StatexIndent}[1][3]{%
  \setlength\@tempdima{\algorithmicindent}%
  \Statex\hskip\dimexpr#1\@tempdima\relax}
\makeatother

\renewcommand{\lim}{\operatornamewithlimits{lim\vphantom{p}}}

\author{Andrea Camisa, Ivano Notarnicola, Giuseppe Notarstefano
\thanks{This result is part of a project that has received funding from the European Research Council (ERC)
    under the European Union's Horizon 2020 research and innovation programme
    (grant agreement No 638992 - OPT4SMART). }    
\thanks{A. Camisa and I. Notarnicola are with the
  Department of Engineering, Universit\`a del Salento, Lecce, Italy, 
  \texttt{name.lastname@unisalento.it.}  }
\thanks{G. Notarstefano is
    with the Department of Electrical, Electronic and Information Engineering, 
    University of Bologna, Bologna, Italy, \texttt{giuseppe.notarstefano@unibo.it.}}
}

\title{A Primal Decomposition Method with Suboptimality Bounds\\ for
  Distributed Mixed-Integer Linear Programming}

\def \algname/{Distributed Primal Decomposition for Feasible MILP Solution}
\def \algacronym/{DiP-FEAS-MILP}

\begin{document}
\maketitle

\begin{abstract}
  In this paper we deal with a network of agents seeking to solve in a
  distributed way Mixed-Integer Linear Programs (MILPs) with a coupling
  constraint (modeling a limited shared resource) and local
  constraints.
  MILPs are NP-hard problems and several challenges arise
  in a distributed framework, so that looking for suboptimal solutions
  is of interest.
  To achieve this goal, the presence of a linear coupling calls for tailored
  decomposition approaches.
  We propose a fully distributed algorithm based on a
  \emph{primal decomposition} approach and a suitable tightening of the coupling
  constraints.
  Agents repeatedly update local allocation vectors, which converge to an
  optimal resource allocation of an approximate version of the original problem.
  Based on such allocation vectors, agents are
  able to (locally) compute a mixed-integer solution, which is guaranteed to be
  feasible after a sufficiently large time.
  Asymptotic and finite-time suboptimality bounds are established for the
  computed solution. Numerical simulations highlight the efficacy of the
  proposed methodology.
\end{abstract}

\section{Introduction}
\label{sec:intro}
In this paper we consider constraint-coupled Mixed-Integer Linear Programs (MILP)
with the following structure
\begin{align}
\label{eq:MILP_original}
\begin{split}
  \min_{\bx_1,\ldots,\bx_N} \: & \: \smallsum_{i =1}^N c_i^\top \bx_i
  \\
  \subj \: 
  & \: \smallsum_{i=1}^N A_i \bx_i \preceq b
  \\
  & \: \bx_i \in X_i, \hspace{1.5cm} i \in \until{N}
\end{split}
\end{align}
where $A_i \in \real^{S \times n_i}$, and $b \in \real^S$
describe $S$ coupling constraints, and the local constraints $X_i$ are
mixed-integer polyhedral, compact and nonempty sets
$X_i = \{\bx_i \in \integer^{Z_i} \times \real^{R_i} \mid D_i \bx_i \preceq d_i \}$,
with $Z_i + R_i = n_i$.
Throughout the paper, we use the symbols $\preceq$ and $\succeq$
to indicate element-wise inequality for vectors.
As customary, we assume~\eqref{eq:MILP_original} is feasible. 
We want to address problem~\eqref{eq:MILP_original} in a distributed computation
framework in which agents of a network communicate only with neighbors.
Since these problems typically arise as instances of dynamic optimization
problems that need to be repeatedly solved in real-time,
we look for \emph{fast} distributed algorithms to compute 
feasible (possibly suboptimal) solutions.
Several problems in cyber-physical network systems as, e.g., in cooperative
robotics or smart grids, can be cast as~\eqref{eq:MILP_original}. An interesting
scenario arises in Distributed Model Predictive Control, where a
large set of dynamical systems, described by local constraints $X_i$
(with both continuous and integer variables), needs to cooperatively solve
a common control task and their states, outputs and/or inputs are
coupled through coupling constraints.

We organize the relevent literature in two parts. First, we review 
primal and dual decomposition methods
for significant types of problems.
In the tutorial papers~\cite{palomar2006tutorial,necoara2011parallel}, primal and
dual decomposition techniques are reviewed.
A distributed primal decomposition approach is proposed in~\cite{lakshmanan2008decentralized}
to solve smooth resource allocation problems.
In~\cite{simonetto2012regularized} a regularized saddle-point algorithm for convex 
networked optimization problems is analyzed.
In \cite{cherukuri2015distributed} distributed algorithms based on
Laplacian-gradient dynamics are used to solve economic dispatch over
digraphs.
In~\cite{nedic2017improved} novel convergence rates for distributed resource 
allocation algorithms are proven.
In~\cite{necoara2013random} the convergence rate of 
a distributed algorithm for the minimization of 
a common cost under a resource constraint is established.
Distributed algorithms for constraint-coupled problems have been
proposed in~\cite{falsone2017dual,notarnicola2017constraint}.
Second, we review parallel and distributed algorithms for
MILPs. In~\cite{kim2013scalable} a Lagrange relaxation approach is used to
decompose MILPs arising in demand response control in smart grids.
In~\cite{takapoui2017simple} a heuristic for embedded mixed-integer programming
is proposed to obtain approximate solutions.
First attempts of proposing a distributed approximate solution for MILPs
are~\cite{kuwata2011cooperative,franceschelli2013gossip}.
Recently, a distributed algorithm, based on cutting-planes,
has been proposed in~\cite{testa2017finite} to solve common cost MILPs.
Although this method could be applied to~\eqref{eq:MILP_original},
each agent would know the entire solution vector (without, e.g.,
preserving privacy).
A pioneering work on fast, master-based parallel algorithms to
find approximate solutions of problem~\eqref{eq:MILP_original}
is~\cite{vujanic2016decomposition}. Here the key idea is to tighten
the coupling constraint and then apply a dual decomposition method to get
mixed-integer points violating the restricted coupling constraint but not the original
one.
In~\cite{falsone2017decentralized}, an improved
\emph{iterative} tightening procedure has been proposed to obtain enhanced
performance guarantees. In both works
\cite{vujanic2016decomposition,falsone2017decentralized}, a master processing
unit is needed.  The first, fully distributed implementation of the above dual
decomposition based methodology is proposed in~\cite{falsone2018phdthesis}.

In this paper we pursue the same main goal as in
\cite{vujanic2016decomposition,falsone2017decentralized,falsone2018phdthesis},
namely to provide a fast algorithm to compute a feasible
mixed-integer solution of \eqref{eq:MILP_original}
with guaranteed suboptimality bounds.
In particular, we consider a \emph{distributed computation framework over
  networks}, even though we will point out how to implement the method in a
parallel architecture.
Differently from the above works, based on dual decomposition, we
propose a distributed algorithm relying on \emph{primal} decomposition.
A building block for the proposed scheme is a distributed algorithm to solve
convex programs. It is based on a relaxation
approach combined with a distributed primal decomposition scheme.
Although the new algorithm strongly relies on the scheme proposed in
\cite{notarnicola2017constraint}, it represents a contribution per se. Indeed,
we give a new insightful interpretation as a primal decomposition scheme
with distributed negotiation among the agents of optimal local allocations.
This distributed algorithm is used to solve a LP approximation
of~\eqref{eq:MILP_original} with restricted coupling constraints.
The resulting (local) allocation vectors allow agents to retrieve a
mixed-integer solution by solving a local MILP with minimal violation
of the allocation vector.
We are able to: (i) establish tight (asymptotic and finite-time)
restrictions of the coupling constraint such that the computed mixed-integer
solutions are feasible for \eqref{eq:MILP_original}, and (ii) provide proper
(asymptotic and finite-time) suboptimality bounds.
Preliminary numerical computations highlight the tightness of our restriction
and show low suboptimality gap.

The paper unfolds as follows. In Section~\ref{sec:set-up} we recall some
preliminaries. In Section~\ref{sec:distributed_algorithm} we derive our
distributed algorithm, analyzed in Section~\ref{sec:analysis}. 
Numerical computations are provided in Section~\ref{sec:simulations}.
Due to space constrains all proofs are omitted and will 
be provided in a forthcoming document.

\section{Preliminaries and Distributed Framework}
\label{sec:set-up}

\subsection{LP Approximation of the Target MILP and
  its Properties}
In order to design fast algorithms to find approximate solutions of
~\eqref{eq:MILP_original}, following \cite{vujanic2016decomposition}, it is
useful to introduce an approximate version of the problem.
The approximation is a linear program obtained by replacing the local
(mixed-integer) constraints $X_i$ with their polyhedral convex hull denoted by
$\convXi$, %
and the total resource vector $b$ with a restricted resource vector
$b - \restriction$, with $\restriction \succeq \0 \in \real^{S}$.
In order to clearly distinguish the decision variables of the original (MILP) and
approximated (LP) problems, from now on, we will use $\bz_i\in\real^{n_i}$ to denote the continuous 
counterpart of the mixed-integer variable $\bx_i \in \integer^{Z_i} \times \real^{R_i}$.
The LP approximation can thus be written as
\begin{align}
\label{eq:LP_restricted}
\begin{split}
  \min_{\bz_1,\ldots,\bz_N} \: & \: \smallsum_{i =1}^N c_i^\top \bz_i
  \\
  \subj \: 
  & \: \smallsum_{i=1}^N A_i \bz_i \preceq b - \restriction
  \\
  & \: \bz_i \in \convXi, \hspace{0.7cm} i \in \until{N}.
\end{split}
\end{align}

We now briefly discuss some properties of this approximation. First of all, for
$\restriction = \0$, problem \eqref{eq:LP_restricted} is a relaxation of
\eqref{eq:MILP_original}.
A well-known property of such a relaxed problem is that its dual problem
coincides with the dual of~\eqref{eq:MILP_original}, see e.g.,
\cite{geoffrion1974lagrangean}.  Since in general MILP~\eqref{eq:MILP_original}
has a duality gap and problem~\eqref{eq:LP_restricted} enjoys strong duality,
then the costs of~\eqref{eq:LP_restricted}, with $\restriction = \0$,
and~\eqref{eq:MILP_original} differ exactly by the MILP duality gap.

The high-level idea motivating the restriction $\restriction$ is to exploit 
problem~\eqref{eq:LP_restricted} to compute mixed-integer points satisfying the local
constraints $X_i$, but violating the coupling constraint
no more than $\restriction$.
We will see in the next section that not only the magnitude, but also the
meaning of the restriction that we use in our approach is different from the
ones proposed in dual decomposition schemes.
A common, standing assumption required for methods based on a given restriction
$\restriction \succeq \0$ of the coupling constraints is the following.
\begin{assumption}[On the restricted LP]%
\label{ass:restricted_LP_feasibility}
  Problem~\eqref{eq:LP_restricted} is feasible and its
  optimal solution is unique.\oprocend
\end{assumption}

Finally, we recall a result in \cite{vujanic2014large} which will play a 
key role in our analysis, shows that any vertex (including the optimal one) 
of~\eqref{eq:LP_restricted} is partially mixed-integer, and provides a bound 
on the number of agents whose solution is not mixed-integer.
\begin{lemma}[{\!\! \cite{vujanic2014large}, Theorem 1}]
\label{lem:LP_integer_components}
Let Assumption~\ref{ass:restricted_LP_feasibility} hold and let
$(\bar{\bz}_1, \ldots, \bar{\bz}_N)$ be a vertex of
problem~\eqref{eq:LP_restricted}. There exists a set
$I_{\INT} \subseteq \until{N}$, with cardinality at least $|I_\INT | \ge N-S-1$,
such that $\bar{\bz}_i \in X_i$ for all $i \in I_{\INT}$.~\oprocend
\end{lemma}

\subsection{Primal Decomposition}
\label{subsec:primal_decomposition}

Primal decomposition allows to obtain a master-subproblem architecture
from constraint-coupled convex programs such as~\eqref{eq:LP_restricted}.
Local \emph{allocation vectors} at each node, adding up to the total
resource $b - \restriction$, are iteratively adjusted until they converge
to the optimal allocation. Thus, each node can asymptotically retrieve
its portion of optimal solution of~\eqref{eq:LP_restricted} by using its local
allocation, \cite{silverman1972primal,bertsekas1999nonlinear}.

 Formally, in a primal decomposition approach, problem~\eqref{eq:LP_restricted}
 can be restated into a hierarchical
 master-subproblem formulation, with a \emph{master problem}
\begin{align}
\begin{split}
  \min_{\by_1,\ldots,\by_N} \: & \: \smallsum_{i =1}^N p_i (\by_i) 
  \\
  & \: \smallsum_{i=1}^N \by_i = b - \restriction
  \\
  & \: \by_i \in Y_i, \hspace{2cm} i \in\until{N}
\end{split}
\label{eq:primal_decomp_master}
\end{align}
where, for each $i\in\until{N}$, $p_i (\by_i) $ is defined as the optimal cost
of the $i$-th \emph{subproblem}
\begin{align}
\begin{split}
  p_i(\by_i) = \: \min_{\bz_i} \: & \: c_i^\top \bz_i
  \\
  \subj \: 
  & \: A_i \bz_i \preceq \by_i
  \\
  & \: \bz_i \in \convXi,
\end{split}
\label{eq:primal_decomp_subproblem}
\end{align}
and $Y_i \subseteq\real^S$ denotes the set of $\by_i$ such that
\eqref{eq:primal_decomp_subproblem} is well-posed (i.e., the set of $\by_i$ such
that there exists at least a $\bar{\bz}_i \in \convXi$ with
$A_i \bar{\bz}_i \preceq \by_i$).

Due to the presence of $Y_i$, solving
\eqref{eq:primal_decomp_master} is not trivial, especially in a distributed
computation framework.
Several works as, e.g.,~\cite{lakshmanan2008decentralized}, investigate a
simplified set-up without local constraints, so that
$Y_i \equiv \real^S$.
Recently, in~\cite{notarnicola2017constraint} a methodology to overcome this
issue has been proposed. We will pursue the same idea to devise a distributed
primal decomposition approach for \eqref{eq:LP_restricted}, that will act as a
building block for our distributed algorithm.

\subsection{Distributed Computation Framework}
\label{sec:distributed_computation_framework}

We consider a network of $N$ processors communicating according to a
\emph{connected} and \emph{undirected} graph $\GG = (\until{N}, \EE)$, where
$\EE\subseteq \until{N} \times \until{N}$ is the set of edges. If $(i,j) \in \EE$, then
nodes $i$ and $j$ can exchange information (and in fact also
$(j,i)\in\EE$). We denote
by $\nbrs_i$
the set of \emph{neighbors} of node $i$ in $\GG$, i.e.,
$\nbrs_i = \left\{j \in \until{N} \mid (i,j) \in \EE \right\}$.
Each node $i$ knows only its local constraint $X_i$, its portion $c_i$ of the
total cost and the matrix $A_i$ of the coupling constraint. 
The goal is that each agent computes an approximation for its portion
$\bx_{i}^\star$ of an optimal solution of~\eqref{eq:MILP_original} by means of
local communication with neighboring agents only.

\section{\algname/}
\label{sec:distributed_algorithm}
In this section we introduce our distributed optimization algorithm and discuss
implementation features.
We provide a constructive argument leading to the proposed method,
consisting of two routines described in the next two subsections.

\subsection{Distributed Primal Decomposition Method for LP Solution over Networks}
Following the approach proposed
in~\cite{notarnicola2017constraint}, we can derive a distributed algorithm to
solve~\eqref{eq:LP_restricted} by combining a (distributed) primal
decomposition method with a relaxation approach.
The algorithm reads as follows. Each agent updates a local vector
$((\bz_i^t, v_i^t), \bmu_i^t)$ as a primal-dual optimal solution pair
of~\eqref{eq:alg_z_LP},
with $M > 0$.
Then, it gathers $ \bmu_{j}^t$ from $j\in\nbrs_i$ and updates its
local estimate of the optimal allocation vector $\by_{i}^{t+1}$
with~\eqref{eq:alg_resource_update},
where $\alpha^t$ is an appropriate step-size sequence.
\begin{assumption}[Diminishing Step-size]
  \label{ass:step-size}
  The step-size sequence $\{ \alpha^t \}_{t\ge0}$, with each $\alpha^t \ge 0$, satisfies the
  conditions $\sum_{t=0}^{\infty} \alpha^t = \infty$,
  $\sum_{t=0}^{\infty} \big( \alpha^t \big)^2 < \infty$.
  \oprocend
\end{assumption}

We can now state the convergence properties of the proposed algorithm,
in which agents solve the relaxed (always feasible) version
\eqref{eq:alg_z_LP} of
problem~\eqref{eq:primal_decomp_subproblem}, and then update their resource
allocation vector $\by_i$ according to a linear update.
\begin{proposition}\label{prop:RSDD_convergence}
  Let Assumptions~\ref{ass:restricted_LP_feasibility} and~\ref{ass:step-size}
  hold and let the local allocation vectors $\by_i^0$ be initialized such 
  that $\sum_{i=1}^N \by_i^0 = b - \restriction$.
  Then, there exists a sufficiently large $M > 0$ for which
  the distributed algorithm \eqref{eq:alg_z_LP},
  \eqref{eq:alg_resource_update} generates an allocation vector sequence
  $\{\by_1^t,\ldots,\by_N^t\}_{t\ge 0}$ such that
  \begin{enumerate}
  \item %
    $\sum_{i=1}^N \by_i^t  = b-\restriction$,
  for all $t\ge 0$, 
   
\item $\lim_{t\to\infty} \| \by_i^t - \by_i^\star \| = 0$ for all
  $i\in\until{N}$, where $(\by_1^\star,\ldots,\by_N^\star)$ is an optimal
  solution of~\eqref{eq:primal_decomp_master};

  \item   
  any limit point of the primal sequence $\{\bz_1^t,\ldots,\bz_N^t\}_{t \ge 0}$ 
  associated to $\{\by_1^t,\ldots,\by_N^t\}_{t\ge 0}$, say $(\bz_1^\infty,\ldots,\bz_N^\infty)$, 
  is an optimal solution of problem~\eqref{eq:LP_restricted},
  and the corresponding cost  
  $\sum_{i=1}^N c_i^\top \bz_{i}^\infty $ is equal to the optimal cost of~\eqref{eq:LP_restricted}.
\oprocend
  \end{enumerate}  
\end{proposition}
\nocite{vanderbeck2005implementing,boyd2004convex,nedic2009approximate,todd1991probabilistic}
As a corollary of \emph{(iii)}, the sequences
$\{v_i^{t}\}_{t\ge 0}$, $i\in\until{N}$, converge to zero.

\subsection{Feasible Mixed-Integer Solution Computation}

The distributed
algorithm~\eqref{eq:alg_z_LP}-\eqref{eq:alg_resource_update},
in general, does not provide asymptotically a mixed-integer solution.
Thus, if $\by_i^\infty$ is the asymptotic assignment of agent $i$, the MILP
\begin{align} 
 \label{eq:x_MILP_description}
      \begin{split}      
        \min_{\bx_{i} } \: &\: c_i^\top \bx_{i}
        \\
        \subj \: & \: A_i \bx_i \preceq \by_i^\infty
        \\
        & \: \bx_i \in X_i,
      \end{split}
\end{align}
admits an optimal solution which is also the solution of
problem~\eqref{eq:primal_decomp_subproblem} with $\by_i = \by_i^\infty$,
for (at least) $N-S-1$ agents.
Thus, if the remaining (at most) $S+1$ agents find a solution to
\eqref{eq:x_MILP_description}, all the agents have a mixed-integer point
satisfying both the local constraints and the coupling constraints.
To this end, it is sufficient that for each agent $i$ for which the LP solution
is not mixed integer, there exists
\emph{at least one} feasible point $\bar{\bx}_i \in X_i$ such that
$A_i \bar{\bx}_i \preceq \by_i^\infty$.

However, this does not happen in general because the negotiated
local allocation vectors are based on local constraints $\convXi$ rather 
than $X_i$.
Thus, we adopt a relaxation approach similar to the one proposed for the
LP approximation. That is, we let agents solve a relaxed version
of~\eqref{eq:x_MILP_description} in which the cost $c_i^\top \bx_{i}$ must be
minimized while allowing for a minimal violation of the coupling constraint.
This can be done by solving~\eqref{eq:alg_lexmin_MILP},
where we use the notation $\lexmin$ to indicate that $\rho_i$ and $\xi_i$
are minimized in a lexicographic order (we will discuss next how to 
solve~\eqref{eq:alg_lexmin_MILP}).

\subsection{Distributed Algorithm: Description and Implementation Discussion}
In the following table we summarize our \algname/ (\algacronym/) algorithm from the
perspective of node $i$.
\begin{algorithm}[H]
\renewcommand{\thealgorithm}{}
\floatname{algorithm}{Distributed Algorithm}

  \begin{algorithmic}[0]

    \Statex \textbf{Initialization}: $\by_{i}^0$ such that $\sum_{i=1}^N \by_i^0 = b - \restriction$ \medskip

    \Statex \textbf{Evolution}: \smallskip
      \StatexIndent[0.75] %
      Compute $\bmu_i^t$ as a dual optimal solution of
      \begin{align} 
      \label{eq:alg_z_LP}
      \begin{split}      
        \min_{\bz_{i}, v_i } \hspace{1.2cm} &\: c_i^\top \bz_{i} + M v_i
        \\
        \subj \hspace{0.3cm} 
        \: \bmu_i : \: & \: A_i \bz_i \preceq \by_i^t + v_i\1
        \\
        & \: \bz_i \in \convXi, \: \: v_i \ge 0
      \end{split}
      \end{align}

      \StatexIndent[0.75] Gather $ \bmu_{j}^t$ from $j\in\nbrs_i$ and update
      \begin{align}
        \label{eq:alg_resource_update}
        \by_{i}^{t+1} = \by_{i}^t + \alpha^t \smallsum_{j \in \nbrs_i} \big( \bmu_{i}^t - \bmu_{j}^t \big)
      \end{align}

      \StatexIndent[0.75] Compute $\bx_i^t$ as an optimal solution of
      \begin{align}
        \label{eq:alg_lexmin_MILP}
        \begin{split}      
          \lexmin_{\rho_i, \xi_i, \bx_i} \: & \: \rho_i
          \\
          \subj \: 
          & \: c_i^\top \bx_i \leq \xi_i
          \\
          & \: A_i \bx_i \preceq  \by_i^t + \rho_i \1
          \\
          & \: \bx_i \in X_i, \: \: \rho_i \ge 0.
        \end{split}
      \end{align}

  \end{algorithmic}
  \caption{\algacronym/}
  \label{alg:algorithm}
\end{algorithm}

\begin{remark}[Parallel Implementation]
We point out that a parallel implementation of the
\algacronym/ distributed algorithm can be obtained by letting a central
unit update the allocation vectors $\by_i^t$. 
This can be done by means of a centralized
subgradient replacing \eqref{eq:alg_resource_update}.\oprocend
\end{remark}

From an implementation point of view, in most cases an explicit description of
$\convXi$ in terms of inequalities might not be available.
Then, column generation techniques can be used to approximate $\convXi$,
see, e.g., \cite{vanderbeck2005implementing}.
However, since the algorithm only requires $\bmu_i^{t+1}$ to evolve, agents can
obtain an estimate by \emph{locally} running a dual subgradient
method to find a dual optimal solution of~\eqref{eq:alg_z_LP} without resorting to a
description of $\convXi$.
Indeed, being the Lagrangian of~\eqref{eq:alg_z_LP} a linear function of the primal variable $\bz_i$, 
a subgradient of the dual function at $\bmu_i^t$ can be easily computed as
$A_i \bar{\bx}_i - \by_i^t$, with $\bar{\bx}_i$ an optimal solution of
\begin{align*}
  \min_{ \bx_i \in X_i } ( c_i^\top + (\bmu_i^t)^\top A_i) \bx_i
  =
  \min_{ \bz_i \in \convXi } ( c_i^\top + (\bmu_i^t)^\top A_i) \bz_i,
\end{align*}
where the equality follows from the linearity of the cost.

Moreover, if agent $i$ wants to know $J_i^{\LP,t} \triangleq c_i^\top \bz_i^t + M v_i^t$ at a given
time instant $t$, by strong duality, can evaluate
$\min_{ \bx_i \in X_i } ( c_i^\top + (\bmu_i^t)^\top A_i) \bx_i -
(\bmu_i^t)^\top \by_i^t$.
The sum of these quantities will appear in the
suboptimality bound in finite-time, so that it is computable in a distributed
way by using an average consensus algorithm.

We now show a simple way to perform the $\lexmin$ optimization
in~\eqref{eq:alg_lexmin_MILP}. First, agents compute $\rho_i^t$ as
the optimal cost of
\begin{align}
  \label{eq:alg_rho_MILP}
  \begin{split}      
    \min_{\rho_i, \bx_i} \: & \: \rho_i
    \\
    \subj \: 
    & \: A_i \bx_i \preceq  \by_i^t + \rho_i \1
    \\
    & \: \bx_i \in X_i, \: \: \rho_i \ge 0.
  \end{split}
\end{align}
Then, they compute $\bx_i^t$ as an optimal solution of
\begin{align}
  \label{eq:alg_x_MILP}
  \begin{split}      
    \min_{\bx_i} \: & \: c_i^\top \bx_i
    \\
    \subj \: & 
    \: A_i \bx_i \preceq  \by_i^t + \rho_i^t \1
    \\
    & \: \bx_i \in X_i.
  \end{split}
\end{align}

\section{Feasibility Guarantees and Suboptimality Bounds of \algacronym/}
\label{sec:analysis}
In this section we analyze the performance of~\algacronym/ distributed
algorithm.
We first derive the restriction on the coupling constraint needed
to ensure asymptotic feasibility.
Then we provide (asymptotic and finite-time) feasibility guarantees and
suboptimality bounds, which can be computed once the algorithm solution is
available.

\subsection{Tight Restriction for Asymptotic Feasibility}

  We derive an upper bound of the violation $\rho_i$ for
  any \emph{admissible} LP allocation $\by_i^\infty$. This allows us to
  find a minimal, a-priori restriction $\restriction$, of the coupling
  constraint.
For all $i$, consider a ``lower bound'' vector $\bL_i$
with components
\begin{align*}
  \bL_i^s = \min_{\bx_i \in X_i} \: A_i^s \bx_i,
\end{align*}
where $A_i^s$ is the $s$-th row of $A_i$. %
$\bL_i^s$ can be equivalently computed by minimizing over $\convXi$.
Thus, each admissible allocation $\by_i^\infty$ satisfies
$\by_i^\infty \succeq \bL_i$.
To compute the maximum violation $\rho_i$,
due to the
mismatch between $\bx_i^t$ and $\bz_i^t$,
let $(\tx_i, \tilde{\ell}_i)$ be an optimal solution to
\begin{align*}
\begin{split}
  \min_{\bx_i, \ell_i} \: & \: \ell_i 
  \\
  \subj \: 
  & \: A_i \bx_i  - \bL_i \preceq \ell_i \1
  \\
  & \: \bx_i \in X_i, \: \ell_i \in \real.
\end{split}
\end{align*}
Then, denoting $\bx_i^\infty$ a solution to
  \eqref{eq:alg_lexmin_MILP} with $\by_i^\infty$, it holds
\[
A_i \bx_i^\infty - \by_i^\infty \preceq A_i \bx_i^\infty - \bL_i
  \preceq \Big(\max_{s\in\until{S}} A_i^s \tx_i - \bL_i^s\Big) \1,
\]
for any admissible $\by_i^\infty$.  
Therefore, a restriction $\restriction$, guaranteeing feasibility of
$(\bx_1^\infty,\ldots,\bx_N^\infty)$ with respect to MILP
\eqref{eq:MILP_original}, can be obtained by setting
$\restriction = \ASYrestriction \1$, with
\begin{align}
  \label{eq:restriction_definition}
  \ASYrestriction \triangleq  
  (S+1) \max_{i\in\until{N}} \max_{s\in\until{S}} \Big( A_i^s \tx_i - \bL_i^s \Big).
\end{align}

\begin{figure}[!htpb]
  \centering
  \hfill
  \begin{tikzpicture}[font=\scriptsize,xscale=0.8,>=stealth]
    \draw[-] (-0.2,0) -- + (0,2.4);
    \draw[->] (-0.4,-0) -- + (4.5,0) 
        node[pos=0,xshift=0.1cm,below] {$0$} 
        node[pos=1,xshift=-0.3cm,below,text width=2cm, align=center] {local resource value};
    
    \def\LW{0.6pt}

    \draw[line width=\LW,draw=gray,fill=gray!10] (1.4,1.9) rectangle +(2, 0.2) 
            node[pos=0,xshift=-0.25cm,yshift=0.3cm] {$\bL_1^1$}
            node[xshift=0.4cm,yshift=0.1cm] {$A_1^1\tx_1$};
    \draw[line width=\LW,draw=BrickRed,fill=BrickRed!10] (1.5,1.7) rectangle + (1.8, 0.2)
            node[pos=0,xshift=-0.25cm,yshift=-0.1cm] {$\bL_2^1$}
            node[xshift=0.4cm,yshift=-0.3cm] {$A_2^1 \tx_2$};

    \draw[line width=\LW,draw=gray,fill=gray!10] (0,1.1) rectangle +(3, 0.2);
    \draw[line width=\LW,draw=BrickRed,fill=BrickRed!10] (0.5,0.9) rectangle + (2.6, 0.2);

    \draw[line width=\LW,draw=gray,fill=gray!10] (0.7,0.3) rectangle + (2.5, 0.2);
    \draw[line width=\LW,draw=BrickRed,fill=BrickRed!10] (0,0.1) rectangle + (2.7, 0.2);
  \end{tikzpicture}
  \hfill
  \begin{tikzpicture}[font=\scriptsize,xscale=0.8,>=stealth]
    \draw[-] (0.,0) -- + (0,2.4);
    \draw[->] (-0.2,-0) -- + (3.8,0) 
        node[pos=0,xshift=0.1cm,below] {$0$} 
        node[pos=1,xshift=-0.2cm,below,text width=1.2cm, align=center] {needed allocation};
    \def\LW{0.5pt}

    \draw[line width=\LW,draw=gray,fill=gray!10] (0,1.9) rectangle +(2, 0.2)
            node[xshift=0.cm,yshift=0.2cm] {$A_1^1\tx_1 - \bL_1^1$};

    \draw[line width=\LW,draw=BrickRed,fill=BrickRed!10] (0,1.7) rectangle + (1.8, 0.2)
            node[xshift=0.cm,yshift=-0.4cm] {$A_2^1\tx_2 - \bL_2^1$};

    \draw[line width=\LW,draw=gray,fill=gray!10] (0,1.1) rectangle +(3, 0.2);
    \draw[line width=\LW,draw=BrickRed,fill=BrickRed!10] (0,0.9) rectangle + (2.6, 0.2);

    \draw[line width=\LW,draw=gray,fill=gray!10] (0,0.3) rectangle + (2.5, 0.2);
    \draw[line width=\LW,draw=BrickRed,fill=BrickRed!10] (0,0.1) rectangle + (2.7, 0.2);

    \draw[line width=0.7pt,densely dotted] (3,0) -- + (0,2.4) 
          node[right] {\small $\frac{\ASYrestriction}{S+1}$};
  \end{tikzpicture}
  \hfill
  \vspace{-0.2cm}
  \caption{Graphical representation of the restriction $\restriction$ for a
    problem with $S=3$ constraints and $N=2$ agents.\vspace{-0.2cm}}
  \label{fig:restriction}
\end{figure}
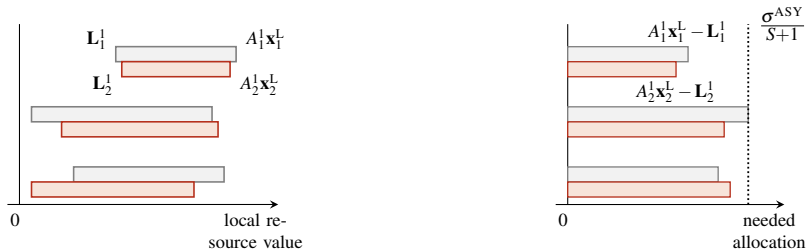

We point out that the restriction~\eqref{eq:restriction_definition} can be
computed in a distributed way by using a max-consensus algorithm.
In Figure~\ref{fig:restriction}, we give an illustrative example of the restriction 
with $3$ constraints and $2$ agents.

Before proceeding with the analysis, we compare our restriction~\eqref{eq:restriction_definition} with
the one proposed in~\cite{vujanic2016decomposition}, i.e.,
\begin{align}
  \label{eq:vujanic restriction}
  \restriction_s = S
  \max_{i \in\until{N}} \Big ( \max_{ \bx_i \in X_i } A_i^s \bx_i - \bL_i^s\Big ),
\end{align}
with $s\in\until{S}$.
Notice that, in place of the worst-case $\max_{ \bx_i \in X_i } A_i^s \bx_i$, we
use $ A_i^s \tx_i$, obtained at the ``lowest'' feasible point
$\tx_i\in X_i$ closest to $\bL_i$.
On the other hand, our scaling factor $S+1$ is larger than $S$.
Finally, we notice that when all the $A_i^s$ are equal for all $s\in\until{S}$, then the two restrictions 
coincide, except for the scaling factor.
This situation occurs for very structured problems, such as in the case study analyzed
in~\cite{vujanic2016decomposition}, for which our restriction
can be obtained by scaling~\eqref{eq:vujanic restriction} by $(S+1)/S$.
 Finally, in certain problem set-ups, such as in partial shipments \cite{vujanic2014large},
  the local constraint sets are such that $\0 \in X_i$ and
  $A_i \bx_i \succeq \0$ for all $\bx_i \in X_i$.
  Then, by construction, $\bL_i = \0$ and $\tilde{\ell}_i = 0$ for all
  $i \in \until{N}$.  It follows that, for these special problems,
  $\ASYrestriction$ boils down to zero, so that no restriction is needed.
  This holds true in general for problems in which $A_i \tx_i = \bL_i$ for all
  $i \in \until{N}$ for some $\tx_i\in X_i$.

\subsection{Asymptotic Guarantees}
\label{sec:asymptotic_guarantees}
In this subsection we assume the allocation vectors
are initialized such that $\sum_{i=1}^N \by_i^0 = b-\restriction$
with $\restriction=\ASYrestriction \1$.

\begin{theorem}[Feasibility]\label{thm:asymptotic_feasibility}
  Let Assumption~\ref{ass:restricted_LP_feasibility} and Assumption~\ref{ass:step-size} hold.
  Consider the allocation vector sequence $\{\by_1^t,\ldots,\by_N^t\}_{t\ge 0}$
  generated by \algacronym/ converging to $(\by_1^\infty, \ldots, \by_N^\infty)$.
  Let $(\rho_i^\infty,\xi_i^\infty,\bx_i^\infty)$ be a lex-optimal solution
  of~\eqref{eq:alg_lexmin_MILP} corresponding to $\by_i^\infty$ for all
  $i\in\until{N}$.
  Then, $(\bx_1^\infty, \ldots, \bx_N^\infty)$  
  is a feasible solution for MILP~\eqref{eq:MILP_original}, i.e.,
  $\bx_i^\infty\in X_i$ for all $i \in \until{N}$ and
  $\sum_{i=1}^N A_i \bx_i^\infty \preceq b$.
  \oprocend
\end{theorem}

Before stating the suboptimality bound, we need 
constraint qualification on the restricted coupling constraint.
\begin{assumption}[Slater Constraint Qualification]\label{ass:slater}
  There exists a point $(\hz_1,\ldots, \hz_N)$ such that $\hz_i \in \conv{X_i}$
  \begin{align}  
    \zeta_{\hz} & \triangleq \min_{s \in \until{S}} \Big( b_s - \restriction_s -\smallsum_{i=1}^N A_i^s \hz_i \Big) > 0 .
    \label{eq:slater_zeta}
  \end{align}
\end{assumption}

The following result establishes an asymptotic suboptimality bound for \algacronym/.
\begin{theorem}[Suboptimality Bound]\label{thm:asymptotic_performance}
  Consider the same assumptions and quantities of Theorem~\ref{thm:asymptotic_feasibility}
  and let also Assumption~\ref{ass:slater} hold.
  Then, $(\bx_1^\infty, \ldots, \bx_N^\infty)$  
  satisfies the suboptimality bound
  \begin{align}
  \label{eq:asymptotic_performance_bound}
  \begin{split}
    \smallsum_{i=1}^N c_i^\top \bx_i^\infty - J^\MILP 
    & \le
    \smallsum_{s=1}^{S+1} ( c_{i_s}^\top \bx_{i_s}^\infty - p_{i_s} ( \by_{i_s}^\infty))
     \\
     & + 
     \dfrac{\ASYrestriction }{\zeta_{\hx}} 
        \smallsum_{i=1}^N \Big( c_i^\top \hx_i 
    - p_i ( \by_i^\infty)
    \Big)
  \end{split}
  \end{align}
  where $J^\MILP$ is the optimal cost of \eqref{eq:MILP_original}, $p_i(\by_i)$ is
  defined in~\eqref{eq:primal_decomp_subproblem}, $(\hz_1,\ldots,\hz_N)$ is a
  Slater point and $i_s$ is the index sequence of the (at most) $S+1$
  agents with $\bz_i^\infty \notin X_i$ (where $\bz_i$ is the local optimal solution
  corresponding to $\by_i^\infty$).%
  \oprocend
\end{theorem}

\subsection{Finite-Time Guarantees}

In this subsection, we establish finite-time guarantees for \algacronym/. 
For the asymptotic results, we used the restriction~\eqref{eq:restriction_definition}.
Here, we consider an augmented restriction $\restriction = (\ASYrestriction + \delta) \1$,
with an arbitrary (small) $\delta > 0$.
\begin{theorem}[Finite-time Feasibility]\label{thm:finite_time_feasibility}
  Let Assumptions \ref{ass:restricted_LP_feasibility} 
  and~\ref{ass:step-size} hold.
  Consider the mixed-integer sequence $\{\bx_1^t,\ldots,\bx_N^t\}_{t\ge 0}$
  generated by \algacronym/
  with $\by_i^0$ initialized such that
  $\sum_{i=1}^N \by_i^0 = b - (\ASYrestriction + \delta) \1$
  for a given $\delta>0$.
  There exists a sufficiently large (finite) time $T_\delta > 0$ such that the
  vector $(\bx_1^t, \ldots, \bx_N^t)$
  is a feasible solution for problem~\eqref{eq:MILP_original}, i.e.,
  $\bx_i^t\in X_i$ for all $i \in \until{N}$ and $\sum_{i=1}^N A_i \bx_i^t \preceq b$,
  for all $t \ge T_\delta$.\oprocend
\end{theorem}
In general, the amount of time $T_\delta$ needed to ensure feasibility increases as
$\delta$ approaches zero.
Conversely, the faster to guarantee feasibility the larger
the restriction should become.
Now, we introduce a suboptimality bound on the computed solution
that can be evaluated when the algorithm is halted prematurely.
\begin{theorem}[Finite-time Suboptimality Bound]\label{thm:finite_time_performance}
  Consider the same assumptions and quantities of Theorem~\ref{thm:finite_time_feasibility}.
  Moreover, let Assumption~\ref{ass:slater} hold and let $\epsilon_i > 0$ be
  arbitrary small numbers, for $i \in \until{N}$.
  Then, there exists a sufficiently large (finite) time $T_\delta > 0$ such that the
  vector $(\bx_1^t, \ldots, \bx_N^t)$ satisfies for all $t\ge T_\delta$ the
  suboptimality bound
  \begin{align}
  \label{eq:finite-time_suboptimality_bound}
  \begin{split}
    \smallsum_{i=1}^N c_i^\top \bx_i^t - J^\MILP 
    & \le
    \smallsum_{i=1}^N ( c_i^\top \bx_i^t - J_i^{\LP,t})%
    \\
    & +
    \smallsum_{i=1}^N \| \bmu_i^t\|_1 \epsilon_i
    +
    \Gamma (\ASYrestriction + \delta),
  \end{split}
  \end{align}
  where $\Gamma = \frac{1}{\zeta_{\hx}} 
    \smallsum_{i=1}^N \Big( \max\limits_{\bx_{i} \in X_i} c_i^\top \bx_i 
    - \min\limits_{\bx_{i} \in X_i} c_i^\top  \bx_{i}
    \Big)$, with $\hz$ a Slater point,
    $J^\MILP$ is the optimal cost of \eqref{eq:MILP_original} and
    $J_i^{\LP,t}$ is the cost of \eqref{eq:alg_z_LP} at time $t$.\oprocend
\end{theorem}

\section{Numerical Computations}
\label{sec:simulations}
In this section, we corroborate the theoretical work with numerical computations.
First, we show that the distributed algorithm achieves feasibility in
finite time (cf. Theorem~\ref{thm:finite_time_feasibility}) on a random MILP
with duality gap.
Second, we compare our approach with methods based on dual decomposition
by generating (unstructured) random MILPs with duality gap.
The generation model for random problems consists of two phases. First, $N$
feasible LPs are generated with a model inspired by \cite{todd1991probabilistic}
to get the local sets $X_i \subset \integer \times \real$ and the cost vectors $c_i$.  
We generate polyhedral constraints $D_i\bx_i \preceq d_i$ with
random entries uniformly in $[0,1]$ for $D_i\in\real^{6\times 2}$ and 
entries in $[0,40]$ for $d_i\in\real^6$. To make sure that $X_i$
is compact, we add box constraints
$-60 \cdot \1 \le \bx_i \le 60 \cdot \1$.
The cost vector is then calculated as $c_i = D_i^\top \hat{c}_i$, where
$\hat{c}_i$ has entries in $[0,5]$. Second, we add coupling constraints by
generating random $A_i$ matrices with entries in $[0,1]$ and a resource
vector $b$ with values in different intervals.

\subsection{Finite-time feasibility}
\begin{figure}[!htpb]
\centering
  \includegraphics[scale=0.90]{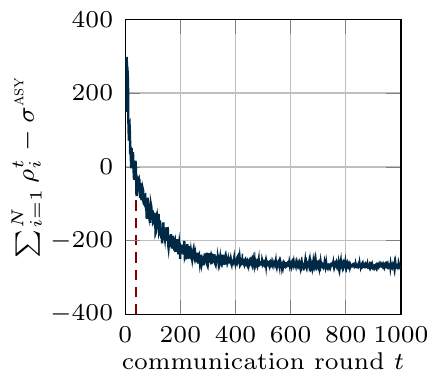}~\hspace{1cm}
  \includegraphics[scale=0.94]{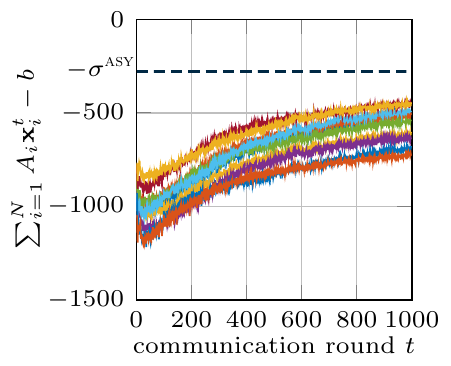}%
  \caption{ Evolution of the sum of local MILP violations compared to the
    restriction $\sigma^{\ASY}$ (left) and coupling constraint value for the
    computed MILP solution (right). \vspace{-0.2cm}}
\label{fig:finite_time_feasibility}
\end{figure}

We generated a random MILP with $N = 100$ agents, $S = 10$ coupling constraints
and resource vector $b$ with entries in $[-600,-500]$.  In order to apply
Theorem~\ref{thm:finite_time_feasibility}, we set the restriction to
$\restriction = (\ASYrestriction + \delta)\1$, with $\delta = 0.4$ and
$\ASYrestriction$ computed with~\eqref{eq:restriction_definition}. 
Remarkably, the solution computed by the algorithm is feasible for the original
MILP for all $t$. In Figure~\ref{fig:finite_time_feasibility} (left), we evaluated the quantity
$\sum_{i=1}^N \rho_i^t - \ASYrestriction$ which in less than $50$ communication rounds
went below $0$, providing a sufficient condition for feasibility
The coupling constraint value is shown in
Figure~\ref{fig:finite_time_feasibility} (right), where the horizontal dashed line
represents $-\ASYrestriction$.  The figure highlights two important
facts: (i) the algorithm seems to evolve in an
interior-point fashion, and (ii) the coupling constraint value is always under
$-\ASYrestriction$, which suggests that there is still room for a tighter
restriction.

\subsection{Performance comparison}
We compared the performance of our algorithm with
\cite{vujanic2016decomposition} whose restriction is defined
in~\eqref{eq:vujanic restriction}.
We did not run the distributed algorithm to compute an optimal allocation
$(\by_1^\infty, \ldots, \by_N^\infty)$, but rather we explicitly solved
problem~\eqref{eq:primal_decomp_master} and then computed
$(\rho_i^\infty, \bx_i^\infty)$ with \eqref{eq:alg_lexmin_MILP}.
We generated problems with $N = 50$ agents and $S = 10$ coupling constraints.

In the numerical computations, we observed that the method in
\cite{vujanic2016decomposition} was not applicable in several generated
instances since the restricted LPs were unfeasible. Therefore, we first
evaluated the fraction of generated problems for which each methodology was
applicable.  It turned out that, for resource vectors with components randomly
chosen in $[-400,-300]$, our methodology could be applied in $99.33\%$ of cases,
out of $300$ instances, whereas the method
\cite{vujanic2016decomposition} never satisfied the needed assumptions (since
the restriction resulted in $0\%$ of feasible restricted LPs).

\begin{figure}[!htpb]
\centering
  \includegraphics[scale=0.9]{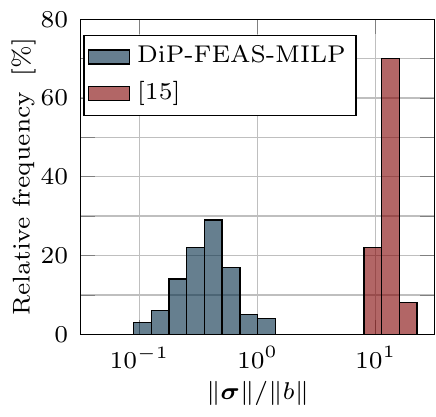}~\hspace{1cm}
  \includegraphics[scale=0.9]{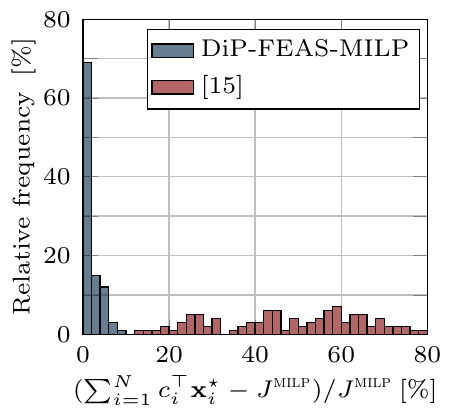}%
\caption{Relative restriction magnitude (left) and solution suboptimality (right) comparison
  for $100$ random MILPs having nonzero duality gap with $S=10$ coupling constraints 
  and $N=50$ agents.\vspace{-0.2cm}}
\label{fig:restriction_comparison}
\end{figure}

Then, we generated problems with entries of $b$ in $[300,400]$ in order
to make both methods applicable.  This made it difficult to find meaningful problems
(i.e., with duality gap), which were only $20.59\%$ out of $2778$ feasible
problems.  For those meaningful problems, we solved the centralized MILP
and we compared the solution performance of \algacronym/ and
\cite{vujanic2016decomposition}. This could be done for $17.66\%$ problems
that were feasible for both methods. In particular, we evaluated the relative
suboptimality $| (\sum_{i=1}^N c_i^\top \bx_i^\star - J^\MILP)/J^\MILP |$, where
$\bx_i^\star$ is the solution found by either \algacronym/ or
\cite{vujanic2016decomposition}. We also evaluated the relative restriction magnitude,
i.e., $\|\restriction\|/\|b\|$.  In
Figure~\ref{fig:restriction_comparison}, comparison histograms of the relative
suboptimality of both methods and of the relative
restriction magnitude are shown.
A further investigation to be carried out consists of comparing the restriction magnitude
of \algacronym/ with the time-varying restriction proposed in \cite{falsone2017decentralized}.

\section{Conclusions}

In this paper we proposed a distributed algorithm for multi-agent MILPs by
employing a primal decomposition approach.  Based on a proper tightening of the
coupling constraints, agents update local allocation vectors that asymptotically
converge to an optimal resource allocation of a convexified version of the
original MILP.  Such vectors allow agents to compute a mixed-integer solution
satisfying both the local constraints and the (original) coupling constraint.
Suboptimality bounds and feasibility guarantees for asymptotic and finite-time
solutions are also established.  Numerical simulations corroborate our approach
showing pros and cons with respect to state-of-the-art methods.

\begin{small}
  \bibliographystyle{IEEEtran}
  \bibliography{primal_decomp_milp_biblio}

\begin{thebibliography}{10}
\providecommand{\url}[1]{#1}
\csname url@samestyle\endcsname
\providecommand{\newblock}{\relax}
\providecommand{\bibinfo}[2]{#2}
\providecommand{\BIBentrySTDinterwordspacing}{\spaceskip=0pt\relax}
\providecommand{\BIBentryALTinterwordstretchfactor}{4}
\providecommand{\BIBentryALTinterwordspacing}{\spaceskip=\fontdimen2\font plus
\BIBentryALTinterwordstretchfactor\fontdimen3\font minus
  \fontdimen4\font\relax}
\providecommand{\BIBforeignlanguage}[2]{{%
\expandafter\ifx\csname l@#1\endcsname\relax
\typeout{** WARNING: IEEEtran.bst: No hyphenation pattern has been}%
\typeout{** loaded for the language `#1'. Using the pattern for}%
\typeout{** the default language instead.}%
\else
\language=\csname l@#1\endcsname
\fi
#2}}
\providecommand{\BIBdecl}{\relax}
\BIBdecl

\bibitem{palomar2006tutorial}
D.~P. Palomar and M.~Chiang, ``A tutorial on decomposition methods for network
  utility maximization,'' \emph{IEEE Journal on Selected Areas in
  Communications}, vol.~24, no.~8, pp. 1439--1451, 2006.

\bibitem{necoara2011parallel}
I.~Necoara, V.~Nedelcu, and I.~Dumitrache, ``Parallel and distributed
  optimization methods for estimation and control in networks,'' \emph{Journal
  of Process Control}, vol.~21, no.~5, pp. 756--766, 2011.

\bibitem{lakshmanan2008decentralized}
H.~Lakshmanan and D.~P. De~Farias, ``Decentralized resource allocation in
  dynamic networks of agents,'' \emph{SIAM Journal on Optimization}, vol.~19,
  no.~2, pp. 911--940, 2008.

\bibitem{simonetto2012regularized}
A.~Simonetto, T.~Keviczky, and M.~Johansson, ``A regularized saddle-point
  algorithm for networked optimization with resource allocation constraints,''
  in \emph{IEEE Conf. on Decision and Control (CDC)}, 2012, pp. 7476--7481.

\bibitem{cherukuri2015distributed}
A.~Cherukuri and J.~Cort{\'e}s, ``Distributed generator coordination for
  initialization and anytime optimization in economic dispatch,'' \emph{IEEE
  Trans. on Control of Network Sys.}, vol.~2, no.~3, pp. 226--237, 2015.

\bibitem{nedic2017improved}
A.~Nedi{\'c}, A.~Olshevsky, and W.~Shi, ``Improved convergence rates for
  distributed resource allocation,'' \emph{preprint arXiv:1706.05441}, 2017.

\bibitem{necoara2013random}
I.~Necoara, ``Random coordinate descent algorithms for multi-agent convex
  optimization over networks,'' \emph{IEEE Trans. on Automatic Control},
  vol.~58, no.~8, pp. 2001--2012, 2013.

\bibitem{falsone2017dual}
A.~Falsone, K.~Margellos, S.~Garatti, and M.~Prandini, ``Dual decomposition for
  multi-agent distributed optimization with coupling constraints,''
  \emph{Automatica}, vol.~84, pp. 149--158, 2017.

\bibitem{notarnicola2017constraint}
I.~Notarnicola and G.~Notarstefano, ``Constraint coupled distributed
  optimization: a relaxation and duality approach,'' \emph{preprint
  arXiv:1711.09221}, 2017.

\bibitem{kim2013scalable}
S.-J. Kim and G.~B. Giannakis, ``Scalable and robust demand response with
  mixed-integer constraints,'' \emph{IEEE Trans. on Smart Grid}, vol.~4, no.~4,
  pp. 2089--2099, 2013.

\bibitem{takapoui2017simple}
R.~Takapoui, N.~Moehle, S.~Boyd, and A.~Bemporad, ``A simple effective
  heuristic for embedded mixed-integer quadratic programming,''
  \emph{International Journal of Control}, pp. 1--11, 2017.

\bibitem{kuwata2011cooperative}
Y.~Kuwata and J.~P. How, ``Cooperative distributed robust trajectory
  optimization using receding horizon {MILP},'' \emph{IEEE Trans. on Control
  Systems Tech.}, vol.~19, no.~2, pp. 423--431, 2011.

\bibitem{franceschelli2013gossip}
M.~Franceschelli, D.~Rosa, C.~Seatzu, and F.~Bullo, ``Gossip algorithms for
  heterogeneous multi-vehicle routing problems,'' \emph{Nonlinear Analysis:
  Hybrid Systems}, vol.~10, pp. 156--174, 2013.

\bibitem{testa2017finite}
A.~Testa, A.~Rucco, and G.~Notarstefano, ``A finite-time cutting plane
  algorithm for distributed mixed integer linear programming,'' in \emph{56th
  IEEE Conf. on Decision and Control (CDC)}, 2017, pp. 3847--3852.

\bibitem{vujanic2016decomposition}
R.~Vujanic, P.~M. Esfahani, P.~J. Goulart, S.~Mari{\'e}thoz, and M.~Morari, ``A
  decomposition method for large scale {MILP}s, with performance guarantees and
  a power system application,'' \emph{Automatica}, vol.~67, no.~5, pp.
  144--156, 2016.

\bibitem{falsone2017decentralized}
A.~Falsone, K.~Margellos, and M.~Prandini, ``A decentralized approach to
  multi-agent {MILP}s: finite-time feasibility and performance guarantees,''
  \emph{preprint arXiv:1706.08788}, 2017.

\bibitem{falsone2018phdthesis}
A.~Falsone, ``Distributed decision making with application to energy systems,''
  \emph{Ph.D.\ dissertation, Politecnico di Milano}, 2018.

\bibitem{geoffrion1974lagrangean}
A.~M. Geoffrion, ``Lagrangean relaxation for integer programming,'' in
  \emph{Mathematical programming study}.\hskip 1em plus 0.5em minus 0.4em\relax
  Springer, 1974, vol.~2, pp. 82--114.

\bibitem{vujanic2014large}
R.~Vujanic, P.~M. Esfahani, P.~Goulart, and M.~Morari, ``Large scale
  mixed-integer optimization: A solution method with supply chain
  applications,'' in \emph{IEEE 22nd Mediterranean Conf. of Control and
  Automation (MED)}, 2014, pp. 804--809.

\bibitem{silverman1972primal}
G.~J. Silverman, ``Primal decomposition of mathematical programs by resource
  allocation: {I}--basic theory and a direction-finding procedure,''
  \emph{Operations Research}, vol.~20, no.~1, pp. 58--74, 1972.

\bibitem{bertsekas1999nonlinear}
D.~P. Bertsekas, \emph{Nonlinear programming}.\hskip 1em plus 0.5em minus
  0.4em\relax Athena scientific, 1999.

\bibitem{vanderbeck2005implementing}
F.~Vanderbeck, ``Implementing mixed integer column generation,'' in
  \emph{Column generation}.\hskip 1em plus 0.5em minus 0.4em\relax Springer,
  2005, pp. 331--358.

\bibitem{boyd2004convex}
S.~Boyd and L.~Vandenberghe, \emph{Convex optimization}.\hskip 1em plus 0.5em
  minus 0.4em\relax Cambridge university press, 2004.

\bibitem{nedic2009approximate}
A.~Nedi{\'c} and A.~Ozdaglar, ``Approximate primal solutions and rate analysis
  for dual subgradient methods,'' \emph{SIAM Journal on Optimization}, vol.~19,
  no.~4, pp. 1757--1780, 2009.

\bibitem{todd1991probabilistic}
M.~J. Todd, ``Probabilistic models for linear programming,'' \emph{Mathematics
  of Operations Research}, vol.~16, no.~4, pp. 671--693, 1991.

\end{thebibliography}
\end{small}

\end{document}